\pdfoutput=1

\documentclass[11pt]{article}

\PassOptionsToPackage{table,dvipsnames}{xcolor}
\usepackage[final]{acl}
\usepackage{times}
\usepackage{latexsym}

\usepackage[T1]{fontenc}

\usepackage[utf8]{inputenc}

\usepackage{microtype}

\usepackage{inconsolata}

\usepackage{graphicx}
\usepackage{algorithm}
\usepackage{algorithmic}
\usepackage{multirow}
\usepackage{amsmath}
\usepackage{amssymb}
\usepackage{amsthm}
\usepackage{booktabs}
\usepackage{tabularx}
\usepackage{appendix}
\usepackage{enumitem}
\usepackage{xspace}
\usepackage{xcolor}
\usepackage{subfigure}
\usepackage{chngcntr}
\usepackage{comment}

\newcommand{\name}{\texttt{HAPMoE}\xspace}%

\title{\name: Heterogeneity-Aware Automatic Parallelism Planning for Mixture-of-Experts Models Training}
\author{
  \textbf{Mengyuan Fan}\textsuperscript{1,2,*},
  \textbf{Peizhuang Cong}\textsuperscript{1,*},
  \textbf{Zixiao Huang}\textsuperscript{3,2,*},
  \textbf{Si Xu}\textsuperscript{2},
  \textbf{Tong Qiao}\textsuperscript{2},
  \\
  \textbf{Yanghao Li}\textsuperscript{2},
  \textbf{Jing Yang}\textsuperscript{2},
  \textbf{Tong Yang}\textsuperscript{1,\textdagger},
  \textbf{Quanlu Zhang}\textsuperscript{2,\textdagger},
  \textbf{Yu Wang}\textsuperscript{3,\textdagger}
  \\
  \\
  \textsuperscript{1}State Key Laboratory of Multimedia Information Processing,\\
  School of Computer Science, Peking University \\
  \textsuperscript{2}Infinigence AI
  \qquad
  \textsuperscript{3}Tsinghua University \\
  \small{\textsuperscript{*}Equal contribution.}
  \quad
  \small{\textsuperscript{\textdagger}Corresponding authors.} \\
  \small{\textbf{Email:}
  \href{mailto:fanmengyuan@stu.pku.edu.cn}
  {fanmengyuan@stu.pku.edu.cn}} \\
  \small{\textbf{Correspondence:}
  \href{mailto:yangtongemail@gmail.com}
  {yangtongemail@gmail.com},
  \href{mailto:zhangquanlu@infini-ai.com}
  {zhangquanlu@infini-ai.com},
  \href{mailto:yu-wang@tsinghua.edu.cn}
  {yu-wang@tsinghua.edu.cn}}
}

\begin{document}
\maketitle
\begin{abstract}
As model sizes continue to scale, distributed training has become inevitable. 
Automatic parallelization techniques can derive efficient training parallelism strategies at low cost while achieving superior performance. 
The difficulty of this problem is jointly determined by the complexity of the model and the underlying compute cluster. 
Meanwhile, mixture-of-experts (MoE) models are increasingly emerging as the dominant architecture and the rapid evolution of accelerator hardware has made cluster heterogeneity commonplace, posing substantial challenges to automatic parallelization. However, existing approaches typically target either MoE architectures or heterogeneous clusters, failing to generalize to scenarios where both challenges coexist.
To this end, we present \name, a heterogeneity-aware automatic parallelism planner for MoE training. \name builds a lightweight MoE-aware cost model and efficiently searches a six-dimensional parallel space, producing parallel plans directly deployable on Megatron-LM. Experiments show that \name improves end-to-end training throughput by up to 3.2× over baselines across heterogeneous clusters. Its non-uniform pipeline partitioning yields an additional up to 78\% gains, and its pruning-enhanced dynamic programming algorithm completes the search within 1 minute, demonstrating high efficiency and practical value in complex hardware environments.
\end{abstract}

\section{Introduction}
As large pretrained models continue to scale, distributed training strategies have become critical for computational efficiency~\cite{naveed2025comprehensive}. 
Manual parallelization based on expert heuristics is inefficient and difficult to optimize for complex models and large search spaces, while automated parallelization search provides a systematic solution with minimal overhead compared to long training runs~\cite{brakel2024model,liang2023survey,chen2024ee}. 
Given the increasing complexity of models and computing resources, efficient and accurate automated parallelization methods have become crucial for large-scale model training.

Two trends make this problem notably harder today.
First, Mixture-of-Experts (MoE) has emerged as a dominant scaling approach by enabling sparse activation and conditional computation~\cite{cai2025survey}. 
However, MoE introduces \emph{dynamic} token routing, heavy All-to-All communication for dispatch/combine, and potentially severe expert load imbalance, complicating accurate performance prediction and stable scaling~\cite{zhou2022mixture,hwang2023tutel}. 
Second, modern training infrastructure is increasingly heterogeneous--mixing accelerators of different generations or vendors, with non-uniform memory capacity and network characteristics--which amplifies pipeline imbalance and makes communication costs highly topology- and device-dependent~\cite{um2024metis}. 
Critically, existing auto-parallel systems rarely address these two trends simultaneously: heterogeneity-aware systems typically assume dense models and do not search MoE-specific dimensions (\textit{e.g.}, EP/TPE), whereas MoE-oriented planners usually assume homogeneous hardware and lack heterogeneity-aware modeling and strategy search. 
As MoE is communication-critical and highly sensitive to bandwidth asymmetry, these assumptions often break down in heterogeneous MoE training.

To this end, we propose \name, a heterogeneity-aware automatic parallelization system for MoE models.
Given an unmodified Megatron-LM training script and a target heterogeneous cluster, \name runs a few warm-up iterations to profile key compute kernels and collective primitives, constructs an MoE-aware cost model that captures sparse expert execution, routing/dispatch behavior, and stage-level memory feasibility, and efficiently searches the 6D parallel space $(DP,PP,TP,CP,EP,TPE)$ to output a deployable parallel plan with the parallel configuration, stage partitioning, device mapping, and recomputation policy.
Experimental results across multiple MoE models and heterogeneous cluster scenarios demonstrate that \name substantially improves training efficiency in real-world production environments.

The main contributions of this paper are as follows:
\begin{itemize}
    \item \textbf{Heterogeneity-aware 6D MoE parallel search.}
    We propose \name, which explicitly covers expert-centric dimensions ($EP/TPE$) on heterogeneous clusters. 
    Across representative mixed-hardware clusters, \name improves end-to-end training throughput by up to \textbf{3.2$\times$}; moreover, in ablations on dense models and non-heterogeneous settings, \name consistently outperforms widely used baselines.
    \item \textbf{Non-uniform PP/DP partitioning for heterogeneous MoE.}
    We enable \emph{non-uniform} pipeline/data-parallel partitioning by jointly optimizing stage-wise parallelism knobs and layer allocation across stages, rather than enforcing uniform $(PP,DP)$ settings. 
    Ablation results show that this design yields \textbf{4\%--78\%} throughput gains depending on cluster composition.

    \item \textbf{Pruning-enhanced DP search with sub-minute overhead.}
    We propose a pruning-enhanced dynamic programming search, which reduces the effective search space from $O(PP \times N^{PP} \times H^{PP})$ to $O(PP \times (N/PP)^{PP} \times H)$ , enabling strategy search to complete within \textbf{$<$1 minute} in all evaluated settings.

\end{itemize}
\section{Background and Related Work}
\subsection{Preliminary}

\textbf{Multi-dimensional Parallelism}. 
Distributed training partitions data, parameters, activations, and computation across devices to overcome single-device compute and memory limits. 
\textbf{Data Parallelism} (\textbf{DP}) replicates parameters and synchronizes gradients via All-Reduce, often with ZeRO for memory efficiency~\cite{rajbhandari2020zero}. 
\textbf{Pipeline Parallelism} (\textbf{PP}) assigns consecutive layers to different devices and overlaps micro-batches, but can suffer from bubbles under stage imbalance~\cite{zheng2022alpa}. 
\textbf{Tensor Parallelism} (\textbf{TP}) shards operators and relies on collectives, performing well on homogeneous high-bandwidth systems but remaining sensitive to heterogeneity~\cite{Tesseract}. 
\textbf{Context Parallelism} (\textbf{CP}) partitions the sequence dimension to reduce activation and KV memory at the cost of extra communication~\cite{jiang2025dcp}. 
For MoE models, \textbf{Expert Parallelism} (\textbf{EP}) distributes experts with token routing and All-to-All exchange, improving scalability while introducing load imbalance~\cite{hwang2023tutel,caishortcut}; \textbf{Tensor-Parallel Experts} (\textbf{TPE}) further shard individual experts to support larger MoE models with higher configuration complexity~\cite{zhang2025comet}.

\textbf{Challenge of MoE Training}.
MoE models scale parameters through sparse activation and dynamic routing, keeping active compute close to that of smaller dense models~\cite{zhou2022mixture,hwang2023tutel}. 
However, routing decisions can drift during training, causing uneven expert loads that are typically mitigated by load-balancing losses and capacity factors~\cite{cong2024prediction}. 
Although sparse activation reduces activation cost, expert parameters and load variation increase memory pressure and out-of-memory risks, especially on heterogeneous devices. 
MoE training is also communication-intensive: dispatch/combine relies on All-to-All primitives and is highly sensitive to bandwidth asymmetry. 
These properties make MoE models scalable, but require resource-aware parallelism that jointly accounts for routing, communication, memory, and hardware heterogeneity.

\subsection{Related Studies}

\textbf{Automatic parallelism}. 
Automatic parallelism has progressed from basic strategy search to multi-dimensional optimization~\cite{liang2023survey}. 
Alpa~\cite{zheng2022alpa} and FlexFlow~\cite{jia2019beyond} explore intra- and inter-operator search using ILP, dynamic programming, or execution simulation, while Unity~\cite{unger2022unity} and AMP~\cite{li2022amp} jointly optimize graph transformations and parallel strategies with cost models. 
DeepSpeed~\cite{rajbhandari2022deepspeed} provides practical heuristic-based 3D parallelism. 
For LLMs, Galvatron~\cite{miao2022galvatron} and Merak~\cite{lai2023merak} integrate multi-dimensional parallelism with pipeline-aware scheduling, and later work improves micro-batch scheduling, synchronization, and throughput prediction via fine-grained simulation~\cite{choi2023towards,li2024automatically}. 
These systems largely target dense models or homogeneous settings, limiting their applicability to heterogeneous MoE training.

\textbf{Distributed training on heterogeneous clusters}. 
Heterogeneous clusters introduce load imbalance and communication inefficiency due to differences in accelerator performance, memory, network, and software stacks. 
HeteroG~\cite{yi2020optimizing} and BytePS~\cite{jiang2020unified} improve utilization through resource-aware scheduling and unified communication. 
Recent systems further incorporate learning-based placement, NIC-aware parallelism, and end-to-end planning, as in HeterPS~\cite{liu2023heterps}, Holmes~\cite{yang2024holmes}, Poplar~\cite{zhang2025poplar}, and Metis~\cite{um2024metis}. 
Other techniques, including asynchronous execution, proportional control, mixed precision, and geo-distributed training~\cite{tyagi2025omnilearn,zong2025training}, improve robustness under dynamic resources. 
Nevertheless, many systems assume specific heterogeneity patterns or dense workloads, and remain brittle under extreme network variability and sparse routing~\cite{strati2025sailor,guo2025cephalo}.

\textbf{Training acceleration for MoE}. 
Large-scale MoE training requires coordinated optimization of routing, expert placement, and cross-device communication. 
MegaScale-MoE~\cite{jin2025megascale} and X-MoE~\cite{yuan2025x} reduce communication overhead through routing and expert-assignment optimizations, while ScMoE~\cite{caishortcut} uses shortcut-connected experts to overlap All-to-All communication with computation.
FasterMoE~\cite{he2022fastermoe} models dynamic expert workloads and optimizes load balancing and communication for large-scale MoE training.
More recent systems target dynamic or heterogeneous training settings: HeterMoE~\cite{wu2025hetermoe} optimizes MoE training on heterogeneous GPUs, while SYMI~\cite{skiadopoulos2026symi} uses adaptive expert replication by decoupling model and optimizer state placement.
These works show the importance of communication- and scheduling-aware MoE optimization, but do not fully integrate heterogeneous hardware modeling with global multi-dimensional parallelism search.

\section{\name Design}

\begin{figure*}[t]
    \centering
    \includegraphics[width=0.9\textwidth]{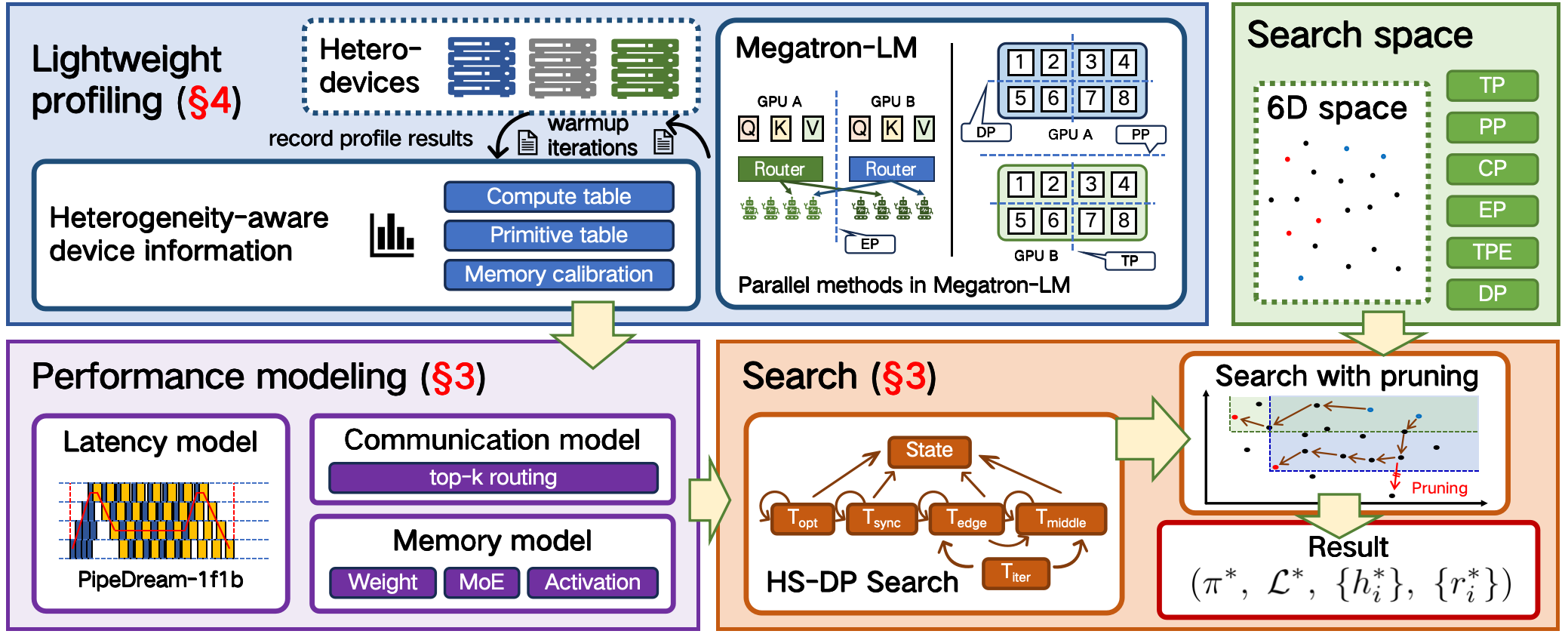}
    \caption{Overview of \name. \name follows a profile-model-search pipeline: it first performs lightweight profiling to build device/primitive lookup tables, then constructs MoE-aware latency/communication/memory models under heterogeneity, and finally searches the 6D parallel space with pruning to output a Megatron-LM-deployable plan.}
    \label{fig:overview}
\end{figure*}

\textbf{Overview}:
\name implements a \emph{profile--model--search} pipeline.
It first runs a few warm-up iterations to profile device kernels and communication primitives, producing lookup tables for compute, memory, and collective latency.
It then builds an MoE-aware cost model that captures sparse expert execution, routing overhead, and dispatch/combine communication under heterogeneous devices.
Finally, \name performs a pruning-enhanced search to jointly decide the 6D parallel configuration $\pi$, a potentially non-uniform pipeline partition $\mathcal{L}$, per-stage device assignment $\{h_i\}$, and recomputation policy $\{r_i\}$.
The resulting plan is directly deployable in Megatron-LM.

\subsection{Problem Formulation}

We consider automatic parallelization of an MoE Transformer with $N$ layers on a heterogeneous cluster.
The cluster is modeled as a device-type set $\mathcal{H}=\{h_1,\ldots,h_{|\mathcal{H}|}\}$, where each type $h$ is profiled into a triple $(P_h,C_h,B_h)$ representing peak compute, memory capacity, and effective bandwidth statistics, respectively.
Each layer may contain an MoE block with $E$ experts and top-$k$ routing.

\noindent\textbf{Input:} model parameters $(N,S,H,E,k)$ and training hyperparameters (batch sizes, optimizer settings, \textit{etc.}); hardware inventory $\{m_h\}$, memory budgets $\{C_h\}$, and profiled bandwidth/collective characteristics.

\noindent\textbf{Output:} a 6D parallel configuration $\pi=(PP,TP,DP,EP,TPE,CP)$, a (possibly) non-uniform pipeline partition $\mathcal{L}=\{n_i\}_{i=1}^{PP}$ with $\sum_i n_i=N$, a per-stage device assignment $\{h_i\}$, and a per-stage recomputation choice $\{r_i\}$.

\noindent\textbf{Objective:} minimize per-iteration latency under per-stage memory constraints:
\begin{equation}
\label{eq:obj}
\begin{split}
&\min_{\pi,\mathcal{L},\{h_i\},\{r_i\}}~ T_{\text{iter}}(\pi,\mathcal{L},\{h_i\},\{r_i\}) \\
&\quad \text{s.t.}\quad M_{\text{stage}}(h_i,n_i,\pi,r_i)\le C_{h_i},~\forall i .
\end{split}
\end{equation}

\subsection{MoE-Aware Performance Modeling}

We estimate stage cost by aggregating per-layer costs and evaluating them through profiled lookup tables.
To keep the model lightweight, most terms are obtained by table lookup; we only expose the MoE-specific scaling and constraints below.

\textbf{Compute and Routing}.
Let $t_{\text{dense}}(h,TP,CP)$ be the profiled time of a dense Transformer layer on device type $h$ under $(TP,CP)$, and let $t_{\text{ffn}}(h,TP,CP)$ be the profiled dense FFN time (replaced by experts).
With expert parallelism $EP$ and optional expert tensor parallelism $TPE$, the per-layer compute time is:
\begin{equation}
\label{eq:moe_comp}
t_{\mathrm{moe}}^{\mathrm{comp}}
\approx
t_{\mathrm{base}}
- t_{\mathrm{ffn}}
+\frac{2\,t_{\mathrm{ffn}}}{\eta_h\cdot EP\cdot TPE}
+ t_{\text{route}}(h),
\end{equation}
where $t_{\mathrm{base}}=t_{\mathrm{dense}}(h,TP,CP)$ and $t_{\mathrm{ffn}}=t_{\mathrm{ffn}}(h,TP,CP)$ are obtained from profiling, and $\eta_h\in(0,1]$ is a calibrated device-efficiency factor.
We model router overhead $t_{\text{route}}(h)$ as a linear term in token count and fit its coefficient during profiling, i.e., $t_{\text{route}}(h)\propto S\cdot B\cdot H$.
PP affects compute only through the stage layer count $n_i$, while DP contributes via $T_{\text{dp-sync}}(\pi)$ and $T_{\text{opt}}(\pi)$ (Eq.~\eqref{eq:iter_latency}).

\textbf{Token Redistribution Communication}.
For top-$k$ routing, the dominant MoE communication is dispatch/combine of token activations across expert groups.
We first estimate the communicated activation volume (bytes):
\begin{equation}
\label{eq:moe_comm_bytes}
D_{\text{moe}}(\pi) ~=~ (S\cdot B)\cdot H\cdot \beta \cdot k \cdot \Bigl(1-\frac{1}{EP}\Bigr),
\end{equation}
where $\beta$ is bytes/element.
Communication time depends on the collective primitive and the link type (intra-node, inter-node, or cross-type).
We therefore use profiled primitive-specific functions
$t_{\text{comm}} \leftarrow f_{\text{prim}}(D,\text{world},\text{link-type})$ and select the best feasible primitive for each candidate.

For native collectives, the overlap realized by the runtime is already reflected in the profiled end-to-end layer time; the separately modeled dispatch/combine term represents the exposed communication cost.

\textbf{Load Imbalance}.
Routing may skew token-to-expert assignment. Let $\mathcal{T}_i$ be tokens assigned to expert $i$.
We quantify imbalance by
\[
\rho_{\text{imb}} = \frac{\max_{i \in [1,E]} |\mathcal{T}_i|}{\frac{1}{E} \sum_{j=1}^{E} |\mathcal{T}_j|}\ge 1,
\]
computed from router statistics (or a conservative default).
We apply it to the MoE critical path as
$t_{\text{moe}} \leftarrow t_{\text{moe}}\cdot \bigl(1+\gamma(\rho_{\text{imb}}-1)\bigr)$ with $\gamma\in[0,1]$.

\name targets stable deployment windows; when routing statistics change persistently, the routing profile can be refreshed and the sub-minute planner rerun.

\textbf{Memory Model (stage-level)}.
For stage $i$, we estimate
$M_{\text{stage}}=M_{\text{w}}+M_{\text{act}}+M_{\text{moe-extra}}$.
Weights (params+grads+optimizer states) are sharded by $(TP,DP,EP,TPE)$, while activation memory scales with stored micro-batches and recomputation:
\[
\begin{aligned}
M_{\text{w}}~\propto~ \frac{n_i}{TP\cdot EP\cdot TPE}\cdot \beta_{\text{opt}}(DP),\\
M_{\text{act}}~\propto~ \frac{M_{\text{store}}\cdot n_i}{TP}\cdot \xi(r_i).
\end{aligned}
\]
$M_{\text{moe-extra}}$ accounts for routing metadata and temporary dispatch buffers, scaled with expert parallelism.
Given $(h_i,n_i,\pi)$, \name selects the minimal $r_i$ satisfying $M_{\text{stage}}(h_i,n_i,\pi,r_i)\le C_{h_i}$.
Here $M_{\text{store}}$ is the number of micro-batches whose activations are kept by the pipeline schedule (e.g., 1F1B), and $\xi(r_i)\in(0,1]$ is the activation-memory reduction factor ($\xi=1$ means no recomputation).

\subsection{Heterogeneity-Aware Device Modeling}
\label{sec:device_model}

The MoE-aware model in \S3.3 estimates \emph{per-layer} compute/communication/memory costs under heterogeneity. 
To evaluate a candidate parallel plan, however, we must further translate these local costs into the \emph{end-to-end per-iteration latency} under the execution schedule.
Therefore, we instantiate an iteration-level latency model on top of the profiled heterogeneous device primitives, so that each candidate plan can be scored by $T_{\text{iter}}$ during search.

Concretely, \name models heterogeneity through three profiled components:
(i) a \textbf{compute table} for $t_{\text{dense}}(\cdot)$ and $t_{\text{ffn}}(\cdot)$ under $(TP,CP)$;
(ii) a \textbf{communication primitive table} for collective time $f_{\text{prim}}(\cdot)$ across link types (intra-node, inter-node, and cross-type);
(iii) a \textbf{memory calibration} that maps the analytical activation/optimizer terms to runtime-measured footprints.
All device- and primitive-specific coefficients are obtained from a few warm-up iterations and cached; during search, we only perform table lookups and simple arithmetic.

\textbf{Latency Model for 1F1B}.
We use the standard PipeDream-style 1F1B schedule, where pipeline stages alternate forward and backward passes over micro-batches; a schematic is provided in Appendix~\ref{sec:app_1f1b}.
Let $t_{f,i}$ and $t_{b,i}$ denote the forward/backward time of one micro-batch on stage $i$.
The iteration latency is decomposed as
\begin{equation}
\label{eq:iter_latency}
\begin{aligned}
T_{\mathrm{iter}}
&= T_{\mathrm{edge}} + T_{\mathrm{middle}}
 + T_{\mathrm{dp\text{-}sync}} + T_{\mathrm{opt}},\\
T_{\mathrm{edge}}
&= \sum_{i=1}^{PP} (t_{f,i}+t_{b,i}),
\end{aligned}
\end{equation}
where $T_{\mathrm{dp\text{-}sync}}$ and $T_{\mathrm{opt}}$ are obtained from profiled collectives and calibrated optimizer kernels. 
The steady-state term $T_{\mathrm{middle}}$ is determined by the bottleneck pipeline stage under the 1F1B schedule; its full expression is provided in Appendix~\ref{sec:app_1f1b}.

\subsection{Search Algorithm}
Given the exponential search space, \name employs a heterogeneity-sensitive dynamic programming (HS-DP) search with pruning.
The key is \textbf{non-uniform layer assignment}: the pipeline partition $\mathcal{L}=\{n_i\}_{i=1}^{PP}$ is allowed to vary across stages, so that stage workloads match the heterogeneous compute capacities $P_{h_i}$, rather than enforcing uniform $N/PP$ layers per stage.
Each candidate is evaluated by the iteration-level latency model in \S\ref{sec:device_model} (Eq.~\eqref{eq:iter_latency}) under memory feasibility constraints.

\textbf{Dynamic Programming Formulation}.
We define $\text{DP}[i, j, m]$ as the minimum predicted iteration latency when the first $j$ layers are assigned to the first $i$ pipeline stages, where $m$ denotes the remaining device-group state after placing stage $i$ (\textit{i.e.}, available counts per device type).
The state transition is:
\begin{equation*}
\begin{split}
\text{DP}[i, j, m] = {}& \min_{\substack{k \in [1, j-1] \\ h \in \mathcal{H} \\ m' \in \mathcal{S}}} \Big\{ \text{DP}[i-1, k, m'] \\
&+ T_{\text{stage}}(h, k+1, j, \pi) \\
&+ T_{\text{comm}}(h, h', k, j) \Big\},
\end{split}
\end{equation*}
where $T_{\text{stage}}$ aggregates per-layer compute/communication on device type $h$ for layers $(k+1..j)$, and $T_{\text{comm}}$ accounts for cross-stage communication between adjacent stages (with $h'$ being the device type chosen for the neighboring stage). Both terms are computed via the profiled MoE and heterogeneity-aware models, and the resulting stage costs are assembled into $T_{\text{iter}}$ by Eq.~\eqref{eq:iter_latency}.
The complete HS-DP algorithm is presented in the appendix.

\textbf{Pruning Strategies}.
The raw search space is $O(PP \times N^{PP} \times H^{PP})$. We apply three prunings:

\noindent\textbf{(1) Load-balance pruning.} 
Let $w_h = P_h / \sum_{h'\in\mathcal{H}} P_{h'}$ be the profiled compute fraction of type $h$.
$
n_i \in \bigl[(1-\delta)\,w_{h_i}N,\; (1+\delta)\,w_{h_i}N\bigr],
$
where $\delta$ (defaulting to 0.2) is a conservative tolerance for modeling noise and communication-induced imbalance.
This reduces the partition enumeration from $O(N^{PP})$ to $O((N/PP)^{PP})$.

\noindent\textbf{(2) Memory Constraint.} We enforce $M_{\text{stage}}(h, n_i, \pi, r_i) \le C_h$ by selecting the minimal feasible recomputation $r_i$; otherwise prune.

\noindent\textbf{(3) Heterogeneous Communication Pruning.} We prefer forming EP groups within homogeneous sub-clusters. For cross-type communication, we prune when:
\[
T_{\text{comm}}^{\text{hetero}} > \eta \cdot \bar{T}_{\text{comp}},
\]
where $\bar{T}_{\text{comp}} = \frac{1}{PP} \sum_{i=1}^{PP} (t_{f,i} + t_{b,i})$ and $\eta=2$ by default.

Combined, these reduce the effective space to $O(PP \times (N/PP)^{PP} \times H)$.

\section{Megatron-LM Framework Integration}
\label{sec:megatron_integration}

\name is designed for \emph{drop-in deployment} on Megatron-LM-style training stacks, requiring no model-code refactoring or manual tuning of distributed knobs.
Overall, \name provides a \emph{script-in, script-out} workflow: given the original training entry script, it automatically derives model/training metadata, performs short warm-up instrumentation, and emits a runnable launcher that reproduces the selected heterogeneous parallel plan.

\paragraph{Configuration extraction.}
\name parses Megatron-LM runtime arguments to recover the minimal metadata needed for deployment, including model shape ($N,H,S$), MoE settings ($E$, top-$k$), batch sizes, and optimizer/runtime options.
These metadata serve as the canonical interface between Megatron-LM and \name, enabling consistent instantiation of profiling, feasibility checks, and plan materialization.

\paragraph{Lightweight runtime instrumentation.}
To avoid intrusive engineering, \name instruments only a few stable hook points in the training loop: (i) iteration-level time breakdown (forward/backward/optimizer), (ii) router statistics for dispatch/combine and imbalance, and (iii) runtime memory footprints.
Instrumentation runs only for a few warm-up iterations and is fully removable for normal training, while producing cached tables consumed by the planner.

\paragraph{Plan materialization.}
Given the optimized plan $(\pi^*, \mathcal{L}^*, \{h_i^*\}, \{r_i^*\})$, \name translates it into Megatron-LM-executable process meshes and flags.
This includes (i) standard knobs for $(PP,TP,DP,EP,TPE,CP)$, (ii) a heterogeneous pipeline specification with explicit non-uniform layer partitions and stage-to-device mapping, and (iii) stage-wise recomputation directives to satisfy memory constraints (Eq.~\eqref{eq:obj}).
All settings are assembled into a standalone launcher script, ensuring reproducible execution without further manual edits.

\begin{table}[t]
\footnotesize
\centering
      \begin{tabular}{clc}\toprule
          \textbf{Cluster Scale}&\textbf{Hardware Configuration}& \textbf{Symbol}\\
          \midrule
          \multirow{3}{*}{16 (Homo.)}& H800 (2$\times$8)& 16-1-h\\
          & 910B (2$\times$8)   &  16-2-h\\
          & MI300X (2$\times$8) &  16-3-h\\
          \cline{2-3}
          \multirow{3}{*}{16 (Heter.)}
          & H800 (1$\times$8)+910B (1$\times$8)& 16-1\\
          & H800 (1$\times$8)+MI300X (1$\times$8)& 16-2\\
          & 910B (1$\times$8)+MI300X (1$\times$8)& 16-3\\
          \cline{2-3}
          \multirow{3}{*}{24}& H800 (1$\times$8)+910B (2$\times$8) & 24-1\\
          & H800 (1$\times$8)+MI300X (2$\times$8)& 24-2\\
          & 910B (2$\times$8)+MI300X (1$\times$8)& 24-3\\
          \cline{2-3}
          \multirow{3}{*}{32}& H800 (2$\times$8)+910B (2$\times$8)& 32-1\\
          & H800 (2$\times$8)+MI300X (2$\times$8)& 32-2\\
          & 910B (2$\times$8)+MI300X (2$\times$8)& 32-3\\ 
          \bottomrule
          \end{tabular}
    \caption{Symbol definitions of clusters}
\label{tab:define}
\end{table}

\begin{figure*}
    \centering
    \includegraphics[width=0.95\linewidth]{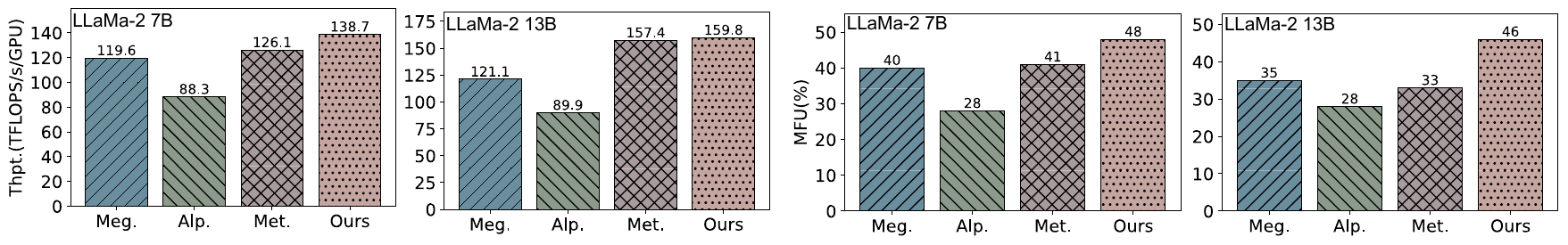}
    \caption{Performance of dense models on heterogeneous clusters: Throughput and MFU.}
    \label{fig:dense-hetero}
\end{figure*}

\begin{figure*}
    \centering
    \includegraphics[width=0.95\linewidth]{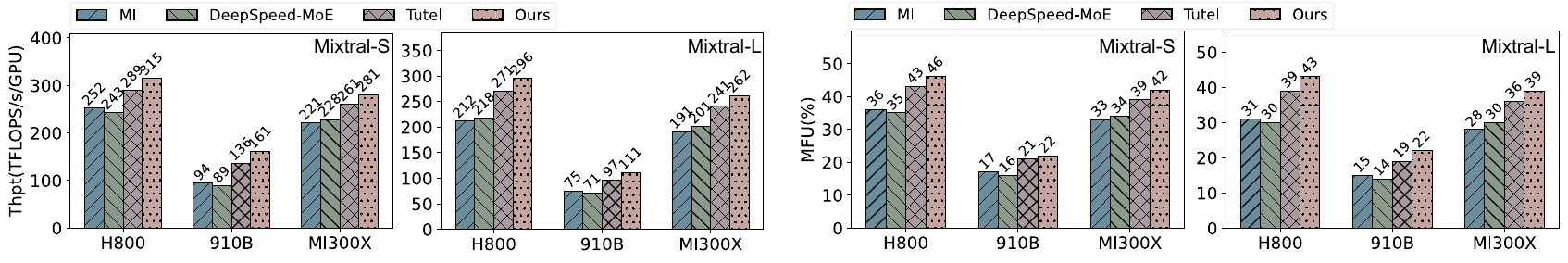}
    \caption{Performance of MoE model on homogeneous clusters: Throughput and MFU.}
    \label{fig:moe-homo}
\end{figure*}

\section{Evaluation}
We evaluate \name through 3 progressive scenarios: dense models on heterogeneous clusters, MoE models on homogeneous clusters, and the core scenario of MoE models on heterogeneous clusters. This systematic comparison benchmarks \name against existing works while highlighting its performance in complex heterogeneous MoE training.

\subsection{Experimental setup}
\textbf{Model}. 
We employ two Mixtral-style MoE configurations, denoted Mixtral-S and Mixtral-L \cite{jiang2024mixtral}, as well as the dense LLaMA-2 7B and 13B models. Detailed MoE configurations are provided in Appendix A.
During training, the micro-batch size is fixed to 1, and the maximum sequence length is set to 4096. 
The global batch size is scaled with the cluster size. 
The system search space spans all six dimensions, enabling fine-grained parallel strategy exploration.

\textbf{Cluster}. Our experiments run on NVIDIA H800, AMD MI300X, and Ascend 910B accelerators. We construct both homogeneous and heterogeneous clusters by combining these devices. Detail specifications and cluster identifiers are listed in Table~\ref{tab:define}.

\textbf{Baselines}. 
For dense models on heterogeneous clusters, we compare \name with Megatron-Infinigence (MI)\footnote{Megatron-Infinigence is a Megatron-LM-based~\cite{shoeybi2019megatron} framework that supports heterogeneous clusters.}, Alpa~\cite{zheng2022alpa}, and Metis~\cite{um2024metis}. 
For MoE models on homogeneous clusters, we compare against MI, DeepSpeed-MoE~\cite{rajbhandari2022deepspeed}, and Tutel~\cite{hwang2023tutel}. 
For MoE models on heterogeneous clusters, we additionally include two adapted baselines: \textbf{Metis-style}, which searches heterogeneous PP/DP/TP/CP placement while fixing EP/TPE from homogeneous profiling, and \textbf{HeterMoE-style}, which applies MoE-layer-level scheduling and asymmetric expert assignment without global 6D search. 
All baselines use same model hyperparameters, precision, router configuration, global batch size, and measurement protocol.

\textbf{Metrics}. (1) \textbf{MFU} (Model FLOPs Utilization) measures hardware efficiency by weighting the peak floating-point operations of devices,  which serves as a primary indicator of hardware efficiency in training clusters.
(2) \textbf{Throughput} is measured in TFLOP/s/device to reflect the overall computational speed of the cluster. 
(3) \textbf{Latency} records the execution time of a single training iteration, which directly evaluates the efficiency of the searched parallel strategies. 
(4) \textbf{Estimation Error} measures the error rate of latency and VRAM usage between \name's estimation and actual values in training, evaluating the reliability of \name in complex hardware settings.

\subsection{Results of Dense Model on Heter. Clusters}
We compared MI, Alpa, Metis, and \name on a heterogeneous cluster with 2×8 A100 and 2×8 910B accelerators, and the results are shown in  Figure~\ref{fig:dense-hetero}. 
For both LLaMA-2 7B and 13B, \name consistently achieves higher throughput and MFU, with gains increasing as model scale grows, owing to its ability to adaptively balance workloads across devices and efficiently explore the partitioning search space. In contrast, MI relies on manual tuning, Alpa assumes homogeneous devices, and Metis uses heuristic partitioning, which limits search space exploration. These indicate that \name can effectively mitigate cross-device imbalances and attain better hardware utilization in heterogeneous clusters.

\subsection{Results of MoE Model on Homo. Clusters}
As shown in Figure~\ref{fig:moe-homo}, the results demonstrate that across various homogeneous clusters and model scales, \name consistently outperforms in throughput and MFU, with gains becoming more pronounced as the model scale increases. This advantage stems from \name’s routing-aware parallel configuration search and cost model, which explicitly accounts for routing-induced imbalance and dispatch costs to better balance expert workloads. In contrast, Megatron-MoE and DeepSpeed-MoE incur high All-to-All communication costs and uneven expert loads under large-scale parallelism, while Tutel lacks support for global parallel strategies because it only focuses on single-layer communication optimization.

\begin{figure}[t]
    \centering
    \includegraphics[width=\linewidth]{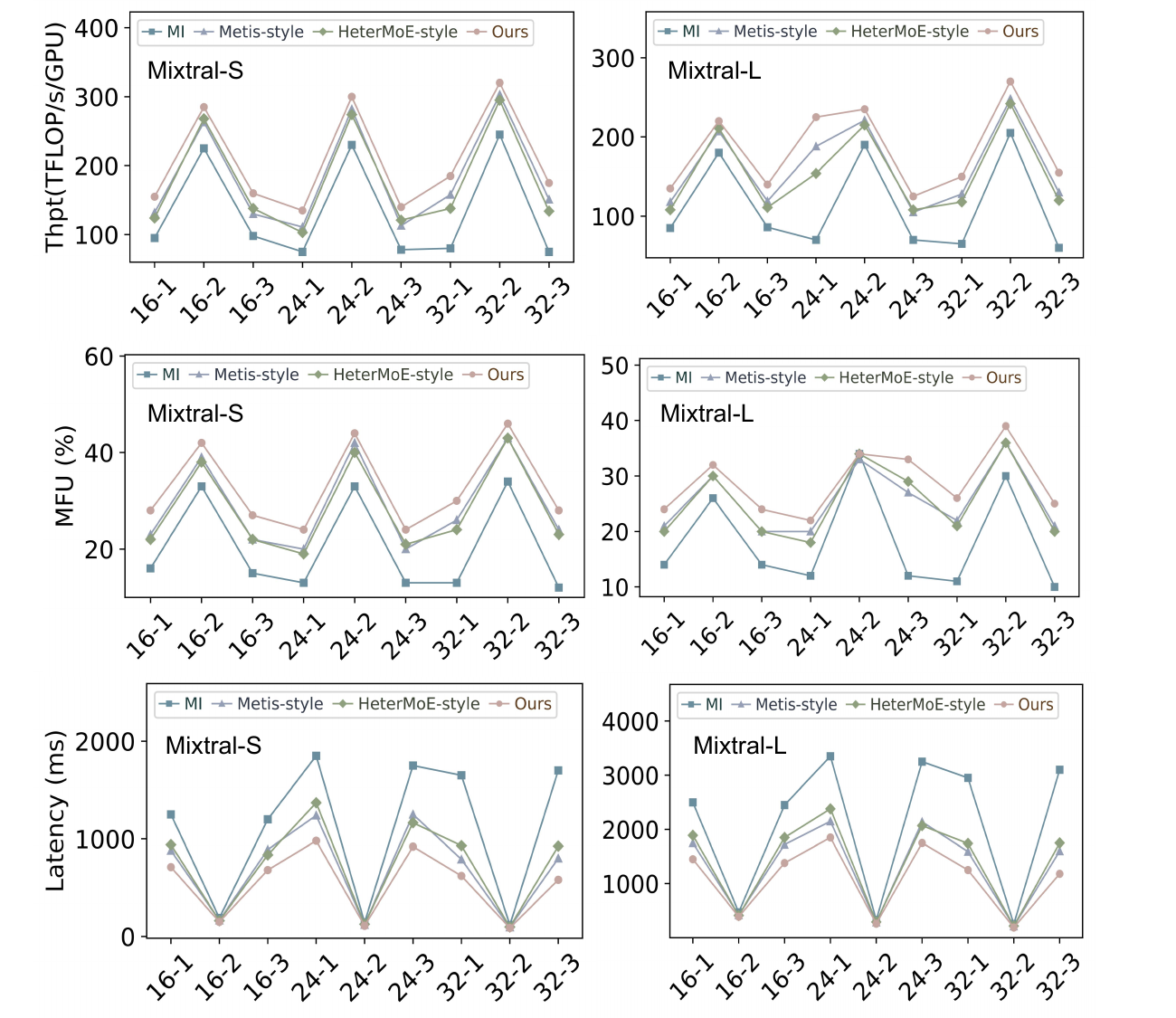}
    \caption{Performance of MoE model on heterogeneous clusters: Throughput, MFU and Latency.}
    \label{fig:moe-heter}
\end{figure}

\subsection{Results of MoE Model on Heter. Clusters}
\label{sec:moe_heter_results}

We conduct end-to-end experiments on Mixtral-S and Mixtral-L across the heterogeneous clusters defined in Table~\ref{tab:define}. 
Besides MI, we include two adapted baselines to better contextualize the results. 

\begin{table}[t]
\footnotesize
\centering
\setlength{\tabcolsep}{4pt}
\begin{tabular}{llccc}
\toprule
\textbf{Model} & \textbf{Method} & \textbf{Thpt.}$\uparrow$ & \textbf{MFU}$\uparrow$ & \textbf{Lat.}$\downarrow$ \\
\midrule
\multirow{4}{*}{Mixtral-S}
& MI & 1.00$\times$ & 1.00$\times$ & 1.00$\times$ \\
& HeterMoE-style & 1.40$\times$ & 1.46$\times$ & 0.72$\times$ \\
& Metis-style & 1.45$\times$ & 1.50$\times$ & 0.69$\times$ \\
& \name & \textbf{1.67$\times$} & \textbf{1.72$\times$} & \textbf{0.56$\times$} \\
\midrule
\multirow{4}{*}{Mixtral-L}
& MI & 1.00$\times$ & 1.00$\times$ & 1.00$\times$ \\
& HeterMoE-style & 1.47$\times$ & 1.50$\times$ & 0.73$\times$ \\
& Metis-style & 1.56$\times$ & 1.53$\times$ & 0.69$\times$ \\
& \name & \textbf{1.78$\times$} & \textbf{1.73$\times$} & \textbf{0.58$\times$} \\
\bottomrule
\end{tabular}
\caption{Geometric mean performance on heterogeneous MoE clusters, normalized to MI. 
Higher is better for throughput and MFU, lower is better for latency.}
\label{tab:moe_heter_norm}
\end{table}

\textbf{Metis-style} extends heterogeneity-aware parallel planning to this setting by searching PP/DP/TP/CP placement while keeping MoE-specific EP/TPE fixed to the best feasible configuration from homogeneous profiling. 
\textbf{HeterMoE-style} applies MoE-layer-level optimizations, including attention/expert disaggregation, overlapped execution, and asymmetric expert assignment, but does not perform global 6D parallel search.

Figure~\ref{fig:moe-heter} reports the per-cluster throughput, MFU, and latency. Table~\ref{tab:moe_heter_norm} summarizes the geometric mean performance normalized to MI. 
Overall, \name achieves the best average performance on both models. 
Compared with MI, \name improves geometric-mean throughput by \textbf{1.67$\times$} on Mixtral-S and \textbf{1.78$\times$} on Mixtral-L, while reducing iteration latency to \textbf{0.56$\times$} and \textbf{0.58$\times$}, respectively. 
The baselines also improve over MI, confirming that both heterogeneous placement and MoE-layer scheduling are beneficial. 
Metis-style is generally stronger due to its global device-aware partitioning, while HeterMoE-style performs competitively on several communication-sensitive cases.

These results suggest that the key advantage of \name comes from jointly considering heterogeneous stage partitioning, device mapping, MoE-specific EP/TPE choices, and memory-feasible recomputation in a unified search space, rather than optimizing any single component in isolation.

\begin{figure}[t]
    \centering
    \includegraphics[width=\linewidth]{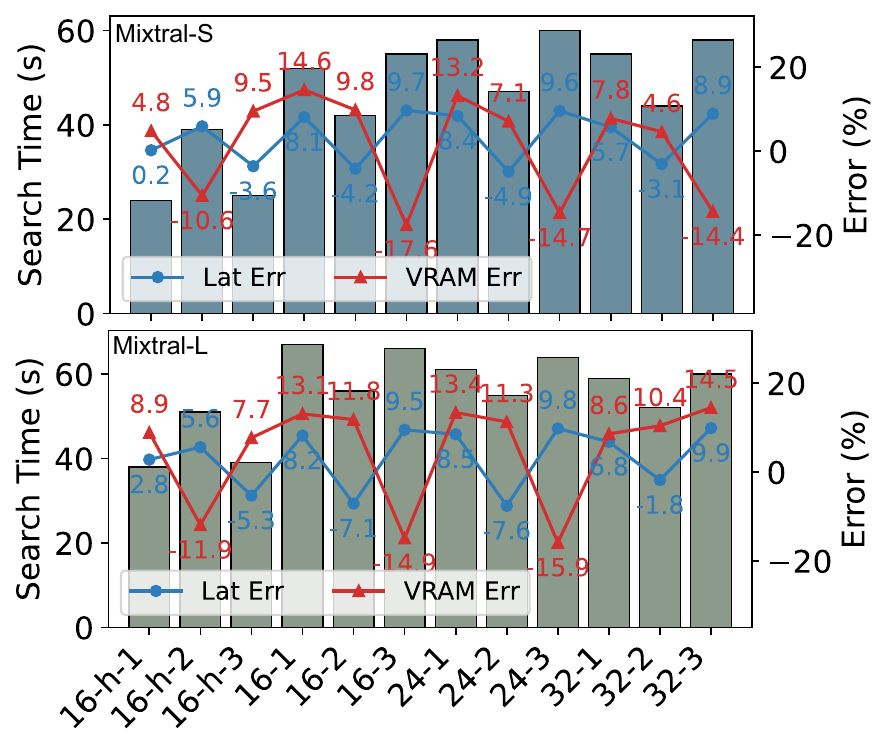}
    \caption{Search time and accuracy of \name}
    \label{fig:searchtime}
\end{figure}

\subsection{Search Time and Accuracy Analysis}
Figure~\ref{fig:searchtime} presents \name’s strategy search time and the estimated accuracy for Latency and VRAM usage across different models and clusters. 
The results show that, regardless of model and cluster scale, search time remains consistently below 60\,s, demonstrating the system’s high search efficiency. 
In terms of accuracy, estimation errors for both metrics remain low, indicating \name’s reliability in performance prediction even under complex configurations. 
Existing studies typically focus on either MoE models or heterogeneous clusters and thus cannot be directly applied to scenarios combining both. 
By providing both low search overhead and high estimation accuracy, \name offers robust support for identifying and searching optimal parallel strategies for complex model training.

\subsection{Disabling Non-uniform PP/DP}
The performance with non-uniform PP/DP disabled (relative to \name) is shown in Table~\ref{tab:ablation_ppdp}. 
On homogeneous clusters, enforcing uniform PP/DP results in only moderate degradation, increasing latency by 10-18\%. 
However, as hardware heterogeneity increases, performance deteriorates sharply: in heterogeneous configurations, latency inflates to 2.3-3.4$\times$, while throughput drops dramatically to 45-62\%. 
Configurations involving 910B are particularly sensitive, as uniform partitioning causes the pipeline steady state to be dominated by the slowest stage, severely constraining overall execution efficiency. 
These results indicate that ignoring device-level performance disparities substantially amplifies pipeline imbalance, underscoring the necessity of non-uniform PP/DP to sustain efficiency on heterogeneous clusters.

\begin{table}[t]
\footnotesize
\centering
\begin{tabular}{lccc}
\toprule
\textbf{Cluster}& \textbf{16-1/2/3-h}& \textbf{16-2} & \textbf{32-1/2/3} \\
\midrule
\textbf{Latency} & 1.12/1.18/1.10$\times$& 2.10$\times$ & 2.30/1.35/2.80$\times$ \\
\textbf{Thpt.}& 0.95/0.93/0.96$\times$& 0.69$\times$ & 0.62/0.85/0.56$\times$ \\
\bottomrule
\end{tabular}
\caption{Performance change ratio by disabling non-uniform PP/DP}
\label{tab:ablation_ppdp}
\end{table}

\paragraph{Profile scaling.}
We further verify that small-scale profiles can guide larger deployments: 4-node profiles predict 16-node latency within about 10\% error across homogeneous and heterogeneous settings, with detailed results in Appendix~\ref{sec:app_profile_scaling}.

\section{Conclusion}
We present \name, a heterogeneity-aware automatic parallelization framework that addresses both MoE sparsity and hardware diversity in large-scale model training. By combining lightweight profiling, an MoE-specific cost model, and an efficient 6-dimensional parallelism search, \name generates optimized hybrid strategies—including non-uniform pipeline and expert parallelism—balancing compute, communication, and memory. Experiments on dense LLMs and MoE models show up to 3.2$\times$ throughput improvement and up to 78\% pipeline gain over SOTA baselines, with minimal search overhead under one minute. \name thus provides a practical, robust solution for efficiently scaling large Transformer models on heterogeneous accelerator clusters.

\clearpage

\clearpage

\section*{Limitations}
HAPMoE currently targets Megatron-LM-style Transformer and MoE training, and builds its plan from warm-up profiles before training starts. While we evaluate multiple heterogeneous accelerator settings, larger production-scale clusters may introduce stronger network contention, failures, and topology effects than those covered in our experiments. HAPMoE also does not yet support online scheduling or dynamic reconfiguration when routing distributions, workload characteristics, or device availability change during training. We leave larger-scale deployment studies and online adaptive scheduling to future work.

\section*{Ethical considerations}
This work studies training-system efficiency and does not introduce new datasets, human-subject data, or user-facing model capabilities. By improving hardware utilization, HAPMoE may reduce the compute cost and energy required for MoE training, while also making large-scale training more accessible. The experiments use existing model architectures and infrastructure measurements.

\section*{Acknowledgments}
This work was supported by the National Key Research and Development Program of China under Grant No.2024YFB2906603, and in part by the National Natural Science Foundation of China (NSFC) under Grant Nos. 62372009 and 62502014.

\bibliography{ref}
\clearpage
\appendix
\renewcommand{\thetable}{\Alph{table}}
\setcounter{table}{0}
\renewcommand{\thefigure}{\Alph{figure}}
\setcounter{figure}{0}

\section*{Appendix}
\section{Model Details}\label{sec:app}
The first model, Mixtral-S ($M_1$), is a medium-scale Mixtral-style MoE Transformer with 24 layers, a hidden size of 4096, an FFN dimension of 14336, and 32 attention heads with grouped-query attention (GQA=8). Its MoE module contains 8 experts with top-$k=2$ routing and uses an all-gather dispatcher. The sequence length is 4096, the micro-batch size is 1, and the global batch size is 256.

The second model, Mixtral-L ($M_2$), is a larger Mixtral-style MoE configuration with 48 layers, a hidden size of 6144, an FFN dimension of 16384, and 48 attention heads, also using grouped-query attention with 8 groups. It contains 8 experts per MoE layer with top-$k=2$ routing and uses an all-to-all dispatcher, comprising approximately 122B parameters. The sequence length is 4096, the micro-batch size is 1, and the global batch size is 512.

Both models use BF16 mixed-precision training with a ZeRO-1 distributed optimizer. For Mixtral-L on the smallest clusters, the FP32 optimizer states are offloaded to host memory.

\section{1F1B Pipeline Schedule Illustration}
\label{sec:app_1f1b}

\begin{figure}[t]
    \centering
    \includegraphics[width=\linewidth]{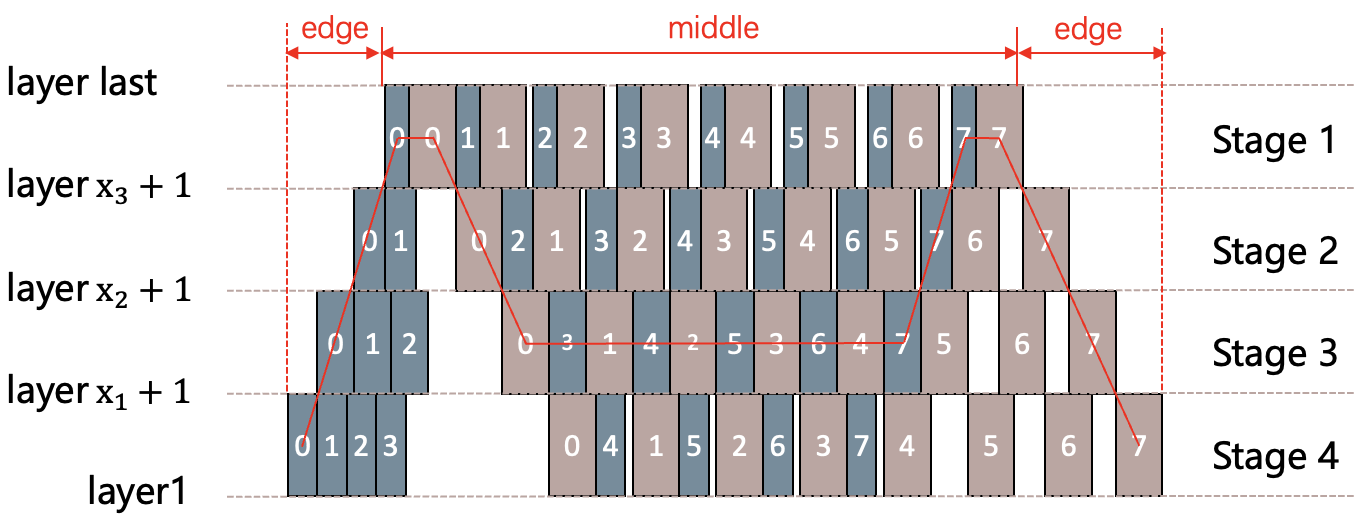}
    \caption{Illustration of the 1F1B pipeline schedule. Forward and backward passes are interleaved across micro-batches, and steady-state throughput is governed by the bottleneck stage.}
    \label{fig:1f1b_schedule_app}
\end{figure}

Figure~\ref{fig:1f1b_schedule_app} illustrates the standard 1F1B schedule used by the iteration-level latency model in Section~\ref{sec:device_model}. The schedule separates warm-up/cool-down steps from the steady state, where throughput is determined by the bottleneck pipeline stage.

Let $s_{\max}=\arg\max_i(t_{f,i}+t_{b,i})$ be the bottleneck stage and $M$ be the number of micro-batches. The steady-state latency term is

\begin{equation}
\begin{aligned}
T_{\mathrm{middle}}
=& \sum_{i=1}^{s_{\max}} t_{f,i} \\
&+ (M-s_{\max}-1)
\max_{1 \leq i \leq PP}
\left(t_{f,i}+t_{b,i}\right).
\end{aligned}
\end{equation}

\section{HS-DP algorithm pseudo-code}
The main text presents the modeling and pruning principles, while Algorithm~\ref{alg:hs_dp} in the appendix lists the complete end-to-end search routine, including stage construction, feasibility filtering, and DP transitions, to ensure reproducibility.

\begin{algorithm}[t]
\caption{Heterogeneity-aware search with pruning (HS-DP).}
\label{alg:hs_dp}
\begin{algorithmic}[1]
\STATE \textbf{Input:} Layers $1..N$, Stages $PP$, Devices $\mathcal{H}$, Configs $\Pi$
\STATE \textbf{Output:} Optimal $(\pi^*, \mathcal{L}^*, \{h_i^*\}, \{r_i^*\})$
\STATE Initialize $\text{DP}[i, j, m] \gets \infty$; $\text{DP}[0, N+1, \cdot] \gets 0$
\STATE Compute $n_{\min,h}$, $n_{\max,h}$ for each $h \in \mathcal{H}$
\FOR{each stage $i = 1$ to $PP$}
    \FOR{each start layer $j = N$ to $1$}
        \FOR{each device $h \in \mathcal{H}$ and end layer $k \in [j+1, N+1]$}
            \STATE $layers \gets k - j$
            \IF{$layers \notin [n_{\min,h}, n_{\max,h}]$}
                \STATE \textbf{continue} \COMMENT{Pruning 1}
            \ENDIF
            \FOR{each config $\pi \in \Pi$}
                \STATE Choose the minimum $r$ such that
                $M_{\text{stage}}(h, layers, \pi, r) \leq C_h$
                \IF{no valid $r$}
                    \STATE \textbf{continue} \COMMENT{Pruning 2}
                \ENDIF
                \IF{$T_{\text{comm}}^{\text{hetero}} > \eta \cdot \bar{T}_{\text{comp}}$}
                    \STATE \textbf{continue} \COMMENT{Pruning 3}
                \ENDIF
                \STATE Compute $T_{\text{stage}}$ and $T_{\text{comm}}$ via the MoE model
                \STATE Update $\text{DP}[i, j, m]$
            \ENDFOR
        \ENDFOR
    \ENDFOR
\ENDFOR
\STATE Reconstruct optimal path via backtrace
\STATE \textbf{return} $(\pi^*, \mathcal{L}^*, \{h_i^*\}, \{r_i^*\})$
\end{algorithmic}
\end{algorithm}

\section{Detailed Results of Profile Scaling}
\label{sec:app_profile_scaling}

To verify that profiling results based on a small-scale cluster can be reliably used for large-scale automatic parallel configuration search, we profile two models on a 4-node cluster and directly reuse the profiling results to guide parallel strategy search on a 16-node cluster. 
In homogeneous clusters, the predicted latency derived from 4-node profiling closely matches the measured latency at 16-node scale, with relative errors consistently bounded within approximately $\pm10\%$ for both models. 
In heterogeneous clusters, the accuracy remains robust despite increased imbalance, with the maximum deviation around 14\%. 
Table~\ref{tab:profile_scaling_details} presents the detailed prediction errors.

\begin{table*}[!t]
\centering
\begin{tabular}{ll c c c c}
\toprule
\multicolumn{2}{c}{\textbf{Cluster}} & \multirow{2}{*}{\textbf{Model}} & \multicolumn{2}{c}{\textbf{Latency} (ms)} & \multirow{2}{*}{\textbf{Error} (\%)} \\
\textbf{Profile}&\textbf{Training} &&\textbf{Pred.}&\textbf{Meas.}&\\
\midrule
H800 (4$\times$8)   &H800 (16$\times$8)   & $M_1$ & 620 & 658 & $-5.78$ \\
MI300X (4$\times$8)   &MI300X (16$\times$8) & $M_1$ & 760 & 718 & $5.85$ \\
H800 (2$\times$8)+MI300X (2$\times$8)&H800 (8$\times$8)+MI300X (8$\times$8)& $M_1$ & 780 & 847 & $-7.91$ \\
H800 (2$\times$8)+910B (2$\times$16) &H800 (4$\times$8)+910B (2$\times$16) & $M_1$ & 1150 & 1336 & $-13.92$ \\
H800 (4$\times$8)   &H800 (16$\times$8)   & $M_2$ & 700 & 637 & $9.89$ \\
MI300X (4$\times$8)   &MI300X (16$\times$8) & $M_2$ & 840 & 897 & $-6.35$ \\
H800 (2$\times$8)+MI300X (2$\times$8)&H800 (8$\times$8)+MI300X (8$\times$8)& $M_2$ & 950 & 862 & $10.21$ \\
\bottomrule
\end{tabular}
\caption{Predicted vs. measured latency of extrapolating 4-node profiling to larger-scale execution}
\label{tab:profile_scaling_details}
\end{table*}

\end{document}